\documentstyle[conf_iap,epsf]{article}
 \def\etal{{\it et~al.\ }}
\def\REFERENCES{\vskip24pt\centerline{\bf REFERENCES}\vskip12pt}
          
          \def\beginrefs{\begingroup\parindent=0pt\frenchspacing 
          \parskip=1pt plus 1pt minus 1pt\interlinepenalty=1000 
          \tolerance=400 \hyphenpenalty=10000
          \everypar={\hangindent=1.6pc}
          
          \def\misref##1,##2,{{\it##1}, {\bf##2},}

          \def\nature##1,{{\it Nature}, {\bf##1},}
          
          \def\aa##1,{{\it Astr.\ Ap.,\ }{\bf##1},}
          \def\aapr{{\it Astr.\ Ap.,\ }in press.}
          \def\ajaa##1,{{\it Astron.\ Astrophys.,\ }{\bf##1},}
          \def\ajaapr{{\it Astron.\ Astrophys.,\ }in press.}
          \def\aalet##1,{{\it Astr.\ Ap.\ (Letters),\ }{\bf##1},}
          \def\aaletpr{{\it Astr.\ Ap.\ (Letters),\ }in press.}
          \def\ajaalet##1,{{\it Astron. Astrophys. (Letters),\ }{\bf##1},}
          \def\ajaaletpr{{\it Astron. Astrophys. (Letters),} in press.}
          
          \def\aasup##1,{{\it Astr. Ap. Suppl.,\ }{\bf##1},}
          \def\aasuppr{{\it Astr.\ Ap.\ Suppl.,\ }in press.}
          \def\ajaasup##1,{{\it Astron. Astrophys. Suppl.,\ }{\bf##1},}
          \def\ajaasuppr{{\it Astron.\ Astrophys.\ Suppl.,\ }in press.}
          
          \def\aass##1,{{\it Astr. Ap. Suppl. Ser.,\ }{\bf##1},}
          \def\aasspr{{\it Astr. Ap. Suppl. Ser.,} in press.}
          
          \def\aj##1,{{\it A.~J.,\ }{\bf##1},}
          \def\ajpr{{\it A.~J.,\ }in press.}
          \def\ajaj##1,{{\it Astron.~J.,\ }{\bf##1},}
          \def\ajajpr{{\it Astron.~J.,} in press.}
          
          \def\apj##1,{{\it Ap.~J.,\  }{\bf##1},}
          \def\apjpr{{\it Ap.~J.,} in press.}
          \def\ajapj##1,{{\it Astrophys.~J.,\ }{\bf##1},}
          \def\ajapjpr{{\it Astrophys.~J.,} in press.}
          
          \def\apjlet##1,{{\it Ap.~J. (Letters),\ }{\bf##1},}
          \def\apjletpr{{\it Ap.~J. (Letters),} in press.}
          \def\ajapjlet##1,{{\it Astrophys. J. Lett.,\ }{\bf##1},}
          \def\ajapjletpr{{\it Astrophys. J. Lett.,} in press.}
           }
          \def\endrefs{\endgroup}
 	
\begin{document}

\heading{  A NEW LOOK AT THE PRIMARY DISTANCE INDICATORS }

\author{D. D. SASSELOV$^1$, J.P. BEAULIEU$^2$}
       {1 Harvard-Smithsonian Center for Astrophysics, Cambridge, 
       MA 02138, USA; dsasselov@cfa.harvard.edu\\
        2 Kapteyn Laboratorium, PO BOX 800, 9700 AV Groningen, Nederlands;
          beaulieu@astro.rug.nl}

\bigskip

\begin{abstract}{\baselineskip 0.4cm 
We examine the overall consistency of the primary distance indicators
and   ways for a  satisfactory resolution of the Cepheid-RR Lyrae zero 
point discrepancy.}
\end{abstract}

\section{Introduction}

The demand on higher precision in distance measurement has prompted us
to examine the systematic uncertainties in the basic primary distance 
indicators. The need for higher {\em accuracy} made us look for overall
consistency. The primary distance indicators are established in our
Galaxy and its companions. Four basic methods provide primary distances
in the near field and as far as Virgo $-$ Cepheids, RR Lyrae, proper
motions, and SN Type II. They agree within large uncertainties
(Huterer, Sasselov, \& Schechter 1995), but
pairwise comparisons of the methods show clearly
a systematic discrepancy between the Cepheids and RR Lyrae, in that the
latter distances are smaller. This discrepancy has been discussed widely
(see Walker 1995 and references therein), and is illustrated in
Fig. 1 $-$ in summary, RR Lyrae calibrations within our Galaxy (BW method,
statistical parallaxes) have a zero point some 0.2$-$0.3~$mag$ fainter.
The RR Lyrae distance scale is the single largest uncertainty in deriving
the ages of globular clusters (Chaboyer \etal 1996), through calibrating
the luminosity of the main sequence turn-off. These ages are often
compared to Hubble timescales, which depend on the Cepheid distance scale.
Clearly any such comparison requires the satisfactory resolution of
the Cepheid$-$RR Lyrae zero point discrepancy.

\section{Cepheids vs. RR Lyrae}

Let us assume that the discrepancy is due to (1) systematic overestimate of
the Cepheid scale, (2) systematic underestimate of the RR Lyr scale, or
(3) combination of the two, and explore some of the possibilities here.

It has been suggested (van den Bergh 1995) that the discrepancy shown
in Fig. 1 might be due to a metallicity-dependent zero point of the Cepheid
PL relation. We note that our calculations illustrated in Fig. 1 already account
for such a metallicity effect, as estimated semi-empirically by Caldwell \&
Coulson (1987) with the use of theoretical model atmospheres. 
However, as theoretical estimates have been uncertain, we
defer our discussion of this idea to the next section, where we report an
empirical result for the metallicity effect on Cepheids.

Most recently, the discrepancy between "high" and "low" RR Lyrae zero points
was strongly reinforced. The MACHO microlensing project published an impressive
sample of double-mode RRd Lyrae variables in the LMC (Alcock \etal 1996). The
distance to LMC can be derived independently (DM = 18.6$\pm0.2$), but the
technique is unfortunately model(s) dependent with systematic uncertainties
which are difficult to estimate. This LMC distance compares well with the one
based on Cepheids (DM = 18.45$\pm0.1$). On the other hand, an equally impressive
sample of galactic RR Lyrae stars was observed and analyzed by Layden \etal
(1996) using the method of statistical parallax. The latter confirms the
"low" zero point and gives DM = 18.28$\pm0.13~mag$ to the LMC.
These two new results are independent of the Cepheid distance scale and
indicate that the problem may lie in part with the RR Lyrae stars themselves.

\begin{figure}
\includegraphics{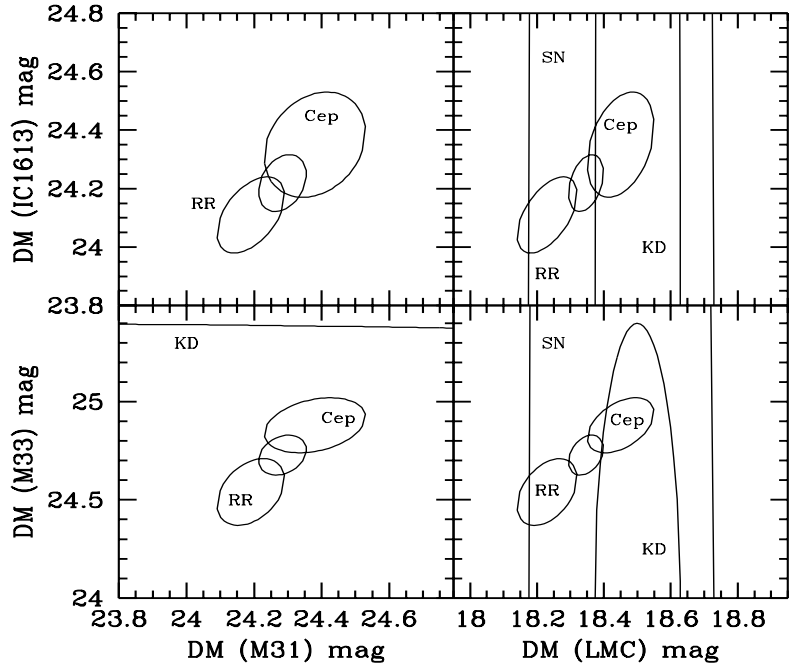}
\vspace*{4.0in}
\caption{Projections of the distance ellipsoids for four distance
methods for four local galaxies  -- the
Large Magellanic Cloud, M31, M33, and IC1613. All these distances are
shown with internal uncertainties, which may be an underestimate in the
case of the RR Lyrae scale. The unmarked ellipse is
the projection of the average distance ellipsoid, based on the simultaneous
solution for 15 objects (see Huterer \etal 1995 for details).}
\end{figure}
\section{Metallicity and the Cepheid zero point}

The degree of dependence of Cepheid distances on chemical
composition (metallicity) has been an open issue for many
years. Here we report on a solution to this problem
from new EROS microlensing survey data. 
EROS (Experience de Recherche d'Objets Sombres) is a collaboration
between French astronomers and particle physicists to search for
baryonic dark matter in the galactic halo through microlensing
effects on stars in the LMC and SMC. We use the CCD data from the
EROS survey to discover about 500 Cepheids and use their $\approx$ 3
million photometric observation for a differential comparison
between LMC and SMC. For more details see Beaulieu \etal (1997) and
Sasselov \etal (1997).

The metallicity effect on the Cepheids from our analysis is:
metal deficiency makes Cepheids bluer, and the period-luminosity
zero point shifts are small and depend on wavelength.
Therefore the metallicity effect on Cepheid distances is large if
the reddening is derived from the colors of the Cepheids.
If one considers as reddening the color difference due to metal
content, $Z$, in deriving the true distance modulus, $\mu$, of a given
galaxy, then we find that the following correction should be applied:
$\delta {\mu}=(0.44_{-0.2}^{+0.1}) ~log\frac{Z}{Z_{LMC}}.$ Such a technique,
which derives
true distance moduli and extinction simultaneously, has been used for M31, M33,
and IC1613. The sign is such that
the distances to metal-poor galaxies would be overestimated if
the effect were ignored $-$ Fig. 2. Obviously, this goes nicely into
solving the discrepancy with the RR Lyrae distance estimate
in IC1613, but not in M31 and M33.
There is much to be desired, however, of the RR Lyrae distances to M31 and M33,
where a modern set of observations is definitely needed (see Huterer \etal 1995
for more details).
\begin{figure}
\includegraphics{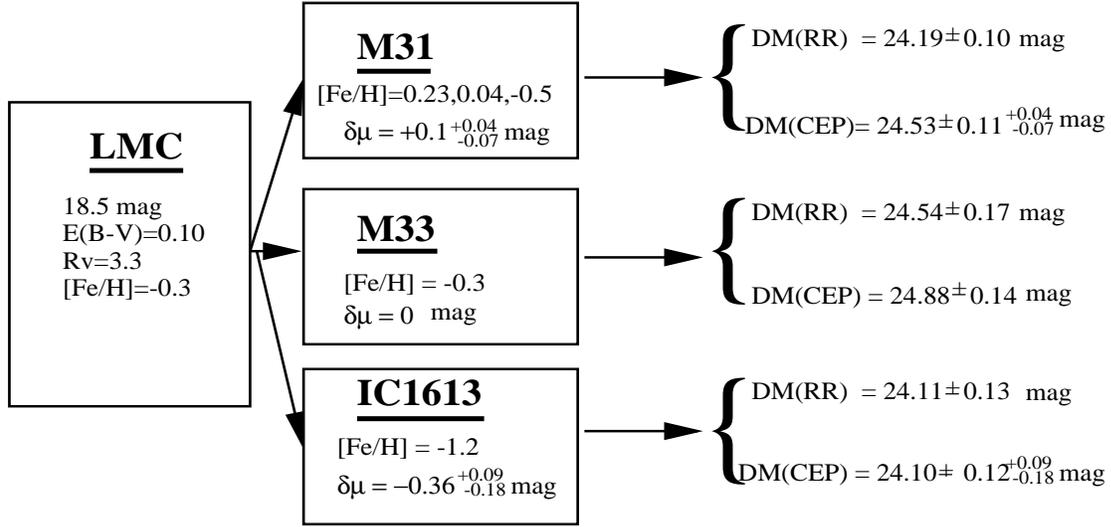}
\vspace*{4.0in}
\caption{RR Lyrae vs. Cepheid distances after correcting for metallicity effects.}
\end{figure}

\section{Conclusions}

Our study of {\bf systematic effects} in the two most heavily weighted
primary distance indicators -- the Cepheids and RR Lyrae, led us
to conclude:\par
 
(1) the Cepheids have a small metallicity dependence, which could
translate to a large correction of the true DM for the standard
distance technique;\par
(2) the Cepheid metallicity dependence is indeed in the right direction
for metal-poor systems, but cannot explain in full the discrepancy
with the RR Lyr scale;\par
(3) hence, the Galactic RR Lyr scale may be underestimated, due to systematics
in the BW and statistical parallax methods (see e.g. Krockenberger, these
proceedings);\par
(4) the corrections due to the above systematics conserve the overall
agreement found for all four primary indicators to 15 objects by
Huterer \etal (1995);\par
(5) finally, the corrections due to the above systematics are (a)
in the direction
of decreasing GCs ages (by $\sim$20\%), if the RR Lyrae zero point is rescaled
to the Cepheids, and (b) converging $H_0$ estimates
to $\approx$70 km.s$^{-1}$Mpc$^{-1}$ due to the metallicity effects on the Cepheid
scale.
 
\REFERENCES

\beginrefs
Alcock, C. et al. 1996, preprint astro-ph/9608036.

Beaulieu, J.P., \etal 1997, \aa, in press.

Chaboyer, B., Demarque, P., Kernan, P.J., \& Krauss, L.M. 1996, Science
271, 957.

Caldwell, J.A.R. \& Coulson, I.M., 1987, \aj 93, 1090.

Huterer, D., Sasselov, D.D., \& Schechter, P.L. 1995, \aj 110, 2705.

Layden, A. C., \etal 1996, preprint astro-ph/9608108.

Sasselov, D.D., \etal 1997, \aa, in press.

van den Bergh, S. 1995, \apj 446, 39.

Walker, A. R. 1995, in {\it Astrophysical Applications of Stellar
Pulsation}, eds. R.S.Stobie and P.A.Whitelock, ASP v.83 (ASP:
San Francisco), p. 198.
\endrefs
\end{document}